\documentstyle [prl,twocolumn,aps,epsfig]{revtex}
\begin{document}
\twocolumn[\hsize\textwidth\columnwidth\hsize\csname@twocolumnfalse%
\endcsname
\title{Magnetic-field-induced superconductivity in layered organic
molecular crystals with localized magnetic moments}
\author{O. C\'epas\cite{email}, Ross H. McKenzie, and J. Merino\cite{address}}
\address{Department of Physics, University of
Queensland, Brisbane, 4072, Australia}

\date{\today}
\maketitle

\begin{abstract}
The synthetic organic compound $\lambda$-(BETS)$_2$FeCl$_4$ undergoes
successive transitions from an antiferromagnetic insulator to a metal
and then to a superconductor as a magnetic field is increased. We use a
Hubbard-Kondo model to clarify the role of the Fe$^{3+}$ magnetic
ions in these phase transitions.  In the high-field regime, the
magnetic field acting on the electron spins is compensated by the
exchange field $H_e$ due to the magnetic ions.  This suggests that the
field-induced superconducting state is the same as the zero-field
superconducting state which occurs under pressure or when the
Fe$^{3+}$ ions are replaced by non-magnetic Ga$^{3+}$ ions.  We show
how $H_e$ can be extracted from the observed splitting of the
Shubnikov-de Haas frequencies.  Furthermore, we use this
method of extracting $H_e$ to predict
the field range for field-induced superconductivity in other
materials.
\end{abstract}

\pacs{PACS numbers: 74.70.Kn, 75.30.Kz}
]

The discovery of magnetic-field-induced superconductivity \cite{uji}
in the two-dimensional compound $\lambda$-(BETS)$_2$FeCl$_4$ [where
BETS is bis(ethylenedithio)-tetraselenafulvalene] is a new example of
the rich phase diagrams of organic molecular crystals \cite{ishiguro}.
Whereas previously pressure or chemical substitution has been used to
change the electronic properties of these organic materials, it is
remarkable that this compound undergoes successive electronic phase
transitions as the magnetic field is increased.  Below a temperature
of 8 K, $\lambda$-(BETS)$_2$FeCl$_4$ is an antiferromagnetic (AF)
insulator\cite{brossard}. As a magnetic field is applied, it undergoes
a first-order transition to a metal at 11 T. Close to this field, the
magnetic moments associated with the spin 5/2 of the Fe$^{3+}$ ions
undergo a transition to a polarized paramagnet.  If the magnetic field
is parallel to the layers, there is a transition to a superconductor
at 20 T\cite{uji}, which is then destroyed above 42
T\cite{brooks}. The magnetic ions are essential to this behavior,
since the compound with non-magnetic ions,
$\lambda$-(BETS)$_2$GaCl$_4$, is, in contrast, a superconductor at
zero field \cite{tanatar}, despite very similar crystal structures
\cite{kobayashi}.

In this Letter we focus on three questions: (i) Why does the inclusion
of magnetic ions change the ground state from a superconductor to an
insulator?  (ii) Is the magnetic-field-induced superconductivity due
to the the Jaccarino-Peter effect\cite{Peter} (as has recently been
proposed \cite{brooks}), where the external field is compensated by an
internal exchange field due to the magnetic ions?  and (iii) Does the
Jaccarino-Peter picture survive if one takes into account the spin
fluctuations associated with the magnetic ions?

Recently, Ziman introduced a two-dimensional Hub\-bard-Kondo model in
order to understand question (i)\cite{brossard}.  The model takes into
account the four conduction bands associated with layers of BETS
molecules (four HOMO orbitals per unit cell), a Kondo coupling between
the localized $S=5/2$ spins and the conduction electrons, and the
Coulomb repulsion between two electrons on the same BETS molecule.
Ziman found that for small electron-electron repulsion the periodic
potential due to the magnetic ordering (found self-consistently) at
low temperature opens energy gaps on the Fermi
surface\cite{brossard}. A magnetic field, by aligning the moments,
destroys the periodic potential, restoring the Fermi-surface.
However, to suppress the entire Fermi-surface, this needs a Kondo
coupling, $J>6$ meV, which is larger than the estimates that we
extract from experiment below.  Moreover, the system seems to have
quite a large electron-electron repulsion, as suggested by comparison
with the $\kappa$-(BEDT-TTF)$_2$X family \cite{mckenzie}.  In this
case, we show first that the system without the magnetic ions may be
close to a Mott transition. Then, the Kondo coupling with the magnetic
ions can drive the system into the insulating phase in order to gain
some magnetic energy.  These two scenarios of the metal-insulator
transition lead to different physical pictures (band-insulator versus
Mott-insulator).

Question (ii) has to be carefully examined.  Although, it is clear
that the magnetic ions can in principle produce an exchange field
$H_e$ that can compensate the external field, it is desirable to know
the precise magnitude of $H_e$.  We show how to extract it from the
observed magnetic oscillations \cite{uji2}.  This allows us to rule
out alternative proposals such as spin-triplet superconductivity,
field-induced dimensional crossovers, or superconductivity mediated by
spin fluctuations in the local moments.

Previous estimates of $J$ involve considerable uncertainty.  In the
high-temperature metallic phase, the exchange leads to an RKKY
interaction between the localized spins, $J_0=J^2 \chi(Q_{AF})$ where
$\chi_{AF}$ is the electronic spin susceptibility at the wave-vector
of the AF correlations. The high-temperature magnetic susceptibility
gives an estimate of $J_{0} \sim 0.2$ meV \cite{brossard}.  To obtain
the coupling $J$ from this approach, we need to know the electronic
spin susceptibility $\chi(Q_{AF})$.  Using the free-electron band
structure, $\chi(Q_{AF})=80 \ \rm (eV)^{-1}$ \cite{brossard} gives
$|J|=1.5 \rm meV$.  Hotta and Fukuyama \cite{hotta} suggested that the
Kondo coupling comes from superexchange processes leading to an
\textit{antiferromagnetic} coupling ($J>0$).  They estimated $J \sim 1
$ meV, using hopping integrals found from H\"uckel calculations and
assuming a value of 2 eV for the splitting between the $d$ orbital of
the Fe$^{3+}$ and HOMO orbitals.

{\it Mott insulator.}  We first argue that the materials without the
magnetic ions are close to a metal-insulator transition. From the
experimental point of view, the effect of the anion in
$\lambda$-(BETS)$_2$GaBr$_z$Cl$_{4-z}$ is to drive the electronic
system from a superconductor for $z<0.8$ to an insulator $z>0.8$
\cite{kobayashidoping}. As the crystal structure is very similar in
both cases, this means that a small change in the electronic
parameters (estimated to be smaller than 5\% \cite{Tanakadoping})
yields two different phases. \textit{ Hence, the electronic system
without magnetic ions is close to a metal-insulator transition}.  From
the theoretical point of view, the $\lambda$-(BETS)$_2$X and
$\kappa$-(BEDT-TTF)$_2$X compounds have very similar band structures:
in these three-quarter filled systems, two bands are isolated from the
two others by quite a large gap \cite{kobayashi}. This can be
interpreted as the separation between the bonding and anti-bonding
orbitals on a dimer of molecules \cite{mckenzie}.  Projecting out the
bonding orbital on each dimer, the system is thus effectively
half-filled and therefore undergoes a metal-Mott insulator transition
if $U/t$ is increased \cite{mckenzie}.  Chemical pressure can change
this ratio driving the system from a metal (or superconductor) to an
insulator \cite{mckenzie}.  Replacing non-magnetic Ga$^{3+}$ by
magnetic Fe$^{3+}$, the electronic parameters change even less
\cite{hotta}.  Even though this could also, in principle, drive the
system from a metal to an insulator, this could not explain why the
metallic phase is restored under magnetic field in a first-order
transition.

We now show that the magnetic character of the ions is important to
drive the system into the insulating phase.  Projecting out
the
bonding-orbitals from Ziman's model leads to a simpler 2-band model,
with Hamiltonian:

\begin{eqnarray}
{\cal H} &=& \sum_{{\bf{ij}},\sigma} t_{\bf ij} (c^{\dagger}_{{\bf{i}},\sigma}
c_{{\bf{j}},\sigma}+h.c.)  +U \sum_{{\bf{i}}} n_{{\bf{i}},\uparrow}
n_{{\bf{i}},\downarrow} + J \sum_{\bf i} \vec{S_{\bf i}}
\cdot \vec{ \sigma_{\bf i}}
\nonumber \\
&+& g_a \mu_B H \sum_{\bf i} S^z_{\bf i}
+ g \mu_B H \sum_{\bf i} \sigma^z_{\bf i}
\nonumber
\end{eqnarray}
where $c^\dagger_{\bf i}$ creates a hole on the dimer at site ${\bf
i}$.  $\vec{S_{\bf i}}$ is a spin-$S$ operator for the local moments.
$\vec{ \sigma_{\bf i}} \equiv {1 \over 2} \sum_{\alpha,\beta} c_{{\bf
i}\alpha}^\dagger \vec{\sigma}_{\alpha \beta} c_{{\bf i},\beta} $
(where $\vec{\sigma}$ denotes the three Pauli matrices) is the
spin-1/2 operator for the hole on site ${\bf i}$.  $U$ and $J$ are,
respectively, the projected Hubbard repulsion and the Kondo coupling.
$t_{\bf ij} $ is the tight-binding hopping integrals between dimers
\cite{mckenzie}. $g_a$ and $g$ are the $g$-factors of the local
moments and itinerant electrons, respectively.

Let us take the two limits of small and large $U$ of this model: (i)
At small $U$ and $J$ small enough, the phase is metallic due to
imperfect nesting \cite{brossard}. The localized spins are subject to
an RKKY interaction.  Treating the local moment spins classically, the
total energy is $E_{metal}-zJ^2 \chi(Q_{AF}) S^2$, where $z=2$ is the
number of magnetic bonds.  (ii) At large $U$, the system is a
Mott-insulator. The electrons are antiferromagnetically ordered
because of the Anderson superexchange process. Subsequently, the Kondo
coupling forces the $S$=5/2 moments to be antiferromagnetically
ordered with respect to the localized electronic spins. The magnetic
energy is $-\frac{1}{2}JS$ per site and the total energy of the AF
Mott insulator (AFMI) is $E_{AFMI}-\frac{1}{2}JS$.  The gain in
magnetic energy is much larger in the Mott phase than in the metallic
phase ($J^2 \chi(Q_{AF}) \sim J^2/E_F \ll J$, where $E_F$ is the Fermi
energy).  Let us assume that the expressions of the magnetic energies
are still valid for intermediate $U$ \cite{caution}.  If for $J=0$,
$E_{AFMI}>E_{metal}$ (the Ga compound is a metal) it is possible that
$E_{AFMI}-\frac{1}{2}JS < E_{Metal}-zJ^2\chi S^2$, provided that $J$
is large enough or the difference between $E_{AFMI}$ and $E_{Metal}$
is small enough.  A similar argument applies to the energy of the
superconducting phase because the RKKY interaction near $Q_{AF}$ is
not modified in the superconducting state \cite{anderson}.

\textit{Destruction of the insulating phase by temperature.}  Above
the Neel ordering temperature ($T_N \sim J_0$) for the local moments
the metallic phase has entropy of order $\ln(2S)$.  In contrast, the
insulating phase with AF order has zero entropy.  Hence, to
zeroth-order in $J_0$, the metal-insulator transition is first order
and occurs at a temperature of $T_{MI} \sim
(E_{metal}(J=0)-E_{AFMI}(J))/\ln(5)$.

\textit{Destruction of the insulating phase by a magnetic field}.  We
calculate the classical energies of the metallic and AFMI states as a
function of the magnetic field. Doing this, we can neglect the
electronic susceptibility because $J_0 \ll t_1, 4 t_1^2/U$.  (i)
\textit{Metallic phase}. We restrict ourselves to spiral ordering such
as $\vec{S}_i=(S\cos \alpha \cos(Q.R_i),S\cos \alpha \sin(Q.R_i),S
\sin \alpha)$. The energy is, $E(H,\alpha)=E_{Metal}-zJ_0S^2\cos 2
\alpha -g_a \mu_B H S \sin \alpha $.  Minimizing this with respect to
$\alpha$ gives: $E(H)=E_{Metal}-zJ_0S^2 -(g_a \mu_B H)^2/8zJ_0$ for
$H<H_N \equiv 4zSJ_0/g_a \mu_B$ which is the the critical field to
align the spins, and $E(H)=E_{Metal}+zJ_0S^2-g_a \mu_B H S$ for
$H>H_N$.  (ii) \textit{Insulating phase}.  The energy is
$E(H,\alpha)=E_{AFMI}-\frac{1}{2}JS\cos \alpha-g_a \mu_B SH \sin
\alpha$.  The minimization gives
$E(H)=E_{AFMI}-\frac{1}{2}JS\sqrt{1+(2g_a \mu_B H/J)^2}$.  Provided
that $E_{Metal} + zJ_0S^2 < E_{AFMI}$, as the field increases the
energy of the metal crosses that of the insulator, leading to a
first-order transition into the metallic phase.

{\it Field-induced superconductivity.}  The argument for the
Jaccarino-Peter mechanism \cite{Peter,brooks} is as follows. If the
system is sufficiently two-dimensional, when a magnetic field is
applied parallel to the layers, the orbital motion of the electrons is
quenched.  The upper critical field is then determined by the Pauli
paramagnetic limit \cite{zuo}.  If we first neglect the fluctuations
of the localized spins and consider the regime where the moments are
aligned by the magnetic field, the Kondo term in the Hamiltonian is
replaced with $J\sum_i \vec{S_i} \cdot \vec{\sigma_i} = - J S \sum_i
\sigma_i^z$. The effective magnetic field experienced by the electrons
is $H-H_e^0$, where $H_e^0=J S /(g\mu_B)$ is a compensating magnetic
field if $J>0$.  At $H=H_e^0$, the Hamiltonian is the same as for the
compound without the magnetic ions ($J=0$) at zero field.  As
$\lambda$-(BETS)$_2$GaCl$_4$ is a superconductor, this mapping shows
that $\lambda$-(BETS)$_2$FeCl$_4$ has to be a superconductor as long
as $|H-H_e^0| < H_P$, the Pauli limiting field. The nature of the
superconductivity in the two materials should therefore be the
same. This is supported experimentally by similar thermodynamic
quantities in both compounds ($T_c^{\rm Ga}=5.5$ K and $T_c^{\rm
Fe}=4.2$ K; $H_{P}^{\rm Ga}=12$ T and $H_{c,max}^{\rm Fe}-H_e \sim 10
$ T).  Tilting of the magnetic field out of plane giving a
perpendicular component of 4 T destroys the superconductivity
\cite{brooks}.  This value is comparable to the upper critical field
for $\lambda$-(BETS)$_2$GaCl$_4$ for perpendicular fields
\cite{tanatar}.  Note also that even if the magnetic field is in the
plane, the orbital limiting field must be larger than $H_e^0$ to get
superconductivity.  This explanation gives $J=1.6 $ meV for $H_e^0=33$
T\cite{brooks}.

{\it Effects of the fluctuations of the localized spins.}  The above
argument neglects the spin flip terms $J \sum_{\bf i} (S^+_{\bf i}
\sigma^-_{\bf i} + S^-_{\bf i} \sigma^+_{\bf i})$ in the Hamiltonian,
where the $+,-$ superscripts denote spin raising and lowering
operators, respectively.  Without the fluctuations, the two
spin-states of the electrons have the same energy for $H=H_e^0$. This
is no longer the case when the spins fluctuate: the spin down can flip
while the spin $S^z=-S$ is raised to $1-S$ at same time. Flipping of
the spin up is, however, blocked because it would require lowering the
spin of the $S_z=-S$ state.  These processes renormalize the
compensating magnetic field. To gain some insight on the relative
importance of this effect, we consider the simple problem of just one
local moment and one electron. The compensating magnetic field is then
given by (when $g \simeq g_a$) $H_e=\frac{4S-1}{4S-2}H_e^0$ (this
reduces to $H_e^0$ for small fluctuations, i.e. large $S$). The real
value of $J$ is therefore slightly larger than that extracted
above. The second effect of the fluctuations
is to increase the on-site repulsion between electrons. Two electrons
on the same site cost not only the energy $U$ but also block the
fluctuations because the spin down is no longer allowed to flip. This
extra repulsion is given by $JS/(4S-2)$, which is negligible compared
to $U$.  In summary, due to the large value of $S$, spin fluctuations
associated with the local moments do not significantly change the
physics.

In order to more clearly establish that the field-induced
superconductivity is due to the compensation effect, it is desirable
to have an independent measurement of the exchange field. We now show
how to extract $H_e$ from the Shubnikov-de Haas oscillations. In
layered organic metals a magnetic field perpendicular to the layers
will produce oscillations in the resistivity that can be related to
the Fermi surface parameters \cite{wosnitza}. In
$\lambda$-(BETS)$_2$FeCl$_4$ at high magnetic field, the magnetic ions
impose an exchange field that splits the conduction bands (for spins
up and down). We calculate the two corresponding frequencies that
should appear in the oscillations \cite{splitting}. In the absence of
an exchange field, as the magnetic field is tilted at an angle
$\theta$ away from the normal to the layers, the oscillatory part is
of the form $\cos[2 \pi F/(H \cos \theta)]$ where $F$ is the
oscillation frequency.  The amplitude of the oscillations is
proportional to the spin splitting factor $R_s = \cos (\pi S_0 /2 \cos
\theta)$, where the argument is proportional to the ratio of the
Zeeman splitting to the Landau level splitting, $S_0=g^* m^*/m_e$,
with renormalized mass and $g$-factor \cite{Shoenberg}. In the
presence of the exchange field, the spin-splitting factor is modified
\cite{Shoenberg1}. We get $ R_s= \cos [\pi S_0 (H_e/H -1)/2 \cos
\theta]$. The effect of this is to produce two oscillation
frequencies, $F/\cos \theta \pm \delta F$ where $\delta F = S_0 H_e/(4
\cos \theta)$.  In $\lambda$-(BETS)$_2$FeCl$_4$, Uji {\it et al.}
observed two frequencies with a difference of 130 T$/\cos \theta$
\cite{uji2}.  If we interpret the frequency difference as due to the
exchange field \cite{extra frequency}, we extract $H_e=32$ T using the
observed effective mass $m^*/m_e = 4.1$, and assuming $g^* =g$
\cite{wilson}. The interpretation of the splitting in terms of a
corrugated three-dimensional Fermi-surface \cite{uji2} is inconsistent
with the observed $1/\cos \theta$ dependence because in that case the
difference should vanish at the Yamaji angles \cite{wosnitza,Yamaji},
seen in the angular-magneto-resistance oscillations \cite{uji2}. Thus
the magnetic oscillations imply that the compensating field should be
about 32 T, in remarkable agreement with the optimal field for
superconductivity.

{\it Electron spin resonance.} It should be pointed out the frequency
splitting discussed above occurs independently of the sign of $J$.
However, the sign of $J$ can be determined unambiguously by ESR. In
the presence of the exchange field, the ESR frequency in the
high-field regime, $\omega = g \mu_B |H-H_e|$ \cite{anisotropy}, will
give $H_e$ and its sign.  Note that, the magnitude of this frequency
should be sufficiently small near $H_e$ that it is experimentally
accessible.

Based on the above picture and the analysis below we predict
field-induced superconductivity in $\kappa$-(BETS)$_2$FeBr$_4$.  It is
an AF metal below 2.5 K, and undergoes a superconducting transition at
1 K \cite{Fujiwara,why}.  The magnetic oscillation spectrum also has
two frequencies with a difference of ${\rm 100 T }/\cos\theta$ and an
effective mass of $m^*/m_e=8$ \cite{Balicas-kBr}. This gives an
exchange field of $|H_e|=12$ T.  The critical field data for
$\kappa$-(BETS)$_2$GaBr$_4$ are not available; but we can estimate
$H_P$ from the critical temperature assuming a BCS relation \cite{zuo}
$H_P^{Ga} \sim 1.8 \ k_B T_c/\mu_B =1.2$ T.  With the above values for
$H_e$ and $H_P$ we would expect field-induced superconductivity in the
range 11 to 13 T if $J>0$.

We now show how the upper critical field parallel to the layers can be
greatly reduced when there is co-existing superconductivity and AF
ordering of the magnetic ions.  This has been dramatically
demonstrated in $\lambda$-(BETS)$_2$FeCl$_4$ under a pressure of 3.5
kbar.  It is an AF metal above 3 kbar \cite{Sato} and undergoes a
superconducting transition at about 1 K \cite{Tanaka_pressure}.
Normally, in layered superconductors the upper critical field parallel
to the layers is much larger than for the field perpendicular to the
layers.  Here, the reverse happens!  The upper critical field parallel
to the layers is only $H_{c2}^\parallel=$ 0.05 T, whereas the
perpendicular critical field is about 0.5 T
\cite{Tanaka_pressure}. This is in contrast with the Pauli limiting
value estimated from the transition temperature, $H_P=2$ T.  We now
show that this rapid destruction of superconductivity by a magnetic
field is due to the polarization of the magnetic ions and it can be
related to the exchange field. In the AF phase, the uniform component
of the spins when a magnetic field is applied is $\langle S_z
\rangle(H,T)$ leading to an exchange field: $J \langle S_z
\rangle(H,T)$. The staggered component of the spins superimposes an
alternating magnetic field that may open gaps in the Fermi-surface. As
the nesting is imperfect, it will not destroy large parts of it and
should not affect strongly the superconducting properties. We will
neglect this effect here. In this case, provided that the crystal
structures of the compounds with and without the magnetic ions are
similar, the upper critical fields of both compounds are related by
$|J \langle S_z \rangle(H_{c2}^{\parallel,Fe}(T),T)-g
\mu_BH_{c2}^{\parallel,Fe}(T)|=g \mu_B H_P^{Ga}(T)$. Measuring the
upper critical fields and the magnetization curve allows a value for
$J$ to be extracted.  For a classical antiferromagnet with exchange
$J_0$, the transverse magnetization is given by $g_a \mu_BH/(4zJ_{0})$
at zero temperature. The relation then becomes $\left|
1-g_a/4zgJ\chi(Q) \right| H_{c2}^{\parallel,Fe}= H_{P}^{Ga}$.  This
shows that $H_{c2}^{\parallel,Fe}$ can be much smaller than $H_P^{Ga}$
(because $J \chi(Q) \sim J/ E_F \ll 1$).

We now apply these ideas to $\kappa$-(BETS)$_2$FeBr$_4$.  The
influence of the magnetic ions has previously been invoked to explain
why the upper critical field is anisotropic within the plane of the
BETS molecules \cite{Fujiwara}.  We rewrite the relation above between
upper critical fields as $|1-H_e/H_N|=
H_P^{Ga}/H_{c2}^{\parallel,Fe}$, having introduced the classical field
to align the moments, $H_N=4zSJ_{0}/g_a \mu_B$ \cite{note1}. This
allows us to extract the parameter $H_e$ (or $J$) from the
measurements of the critical fields. In $\kappa$-(BETS)$_2$FeBr$_4$,
$H_{c2}^{\parallel,Fe} \sim 1$ T for $H\parallel c$, $H_N \sim 5$ T
\cite{Fujiwara} and $H_P^{Ga} \sim 1.2 $ T (see above). The positive
solution is $H_e \sim 10 $ T, consistent with the estimate above.

In conclusion, we have stressed the possibility of having a
\textit{Mott} insulator in $\lambda$-(BETS)$_2$FeCl$_4$ at zero
magnetic field. The measurement of the charge gap as a function of
field may help distinguish the Mott versus band insulator: for the
Mott picture the gap should not vary significantly with field whereas
for the band picture it should. Furthermore, we have shown that the
Hamiltonian that describes $\lambda$-(BETS)$_2$FeCl$_4$ at high fields
is simply related to that for $\lambda$-(BETS)$_2$GaCl$_4$ with a
compensating magnetic field acting on the spins.  We have interpreted
the splitting of the magnetic oscillations as a signature of the
exchange field, thus allowing us to extract the Kondo coupling.  The
strength of the exchange field equals that of the optimal field at
which superconductivity is observed. This strongly supports the
Jaccarino-Peter effect and suggests that the nature of the
superconductivity is the same in both materials.  Note that this
picture is true only if $J>0$ and can be confirmed by ESR.  Using the
same procedure, we have predicted that the organic superconductor
$\kappa$-(BETS)$_2$FeBr$_4$ should also exhibit a field-induced
superconducting phase at about 10T, having extracted the Kondo
coupling both from the magnetic oscillation splitting and from upper
critical field data.

We thank L. Balicas and J.S. Brooks for helpful discussions and for
providing experimental data, prior to publication.  We thank
P.W. Anderson, S. Brown and T. Ziman for helpful discussions. RHM
thanks the Aspen Center for Physics for hospitality.  This work was
supported by the Australian Research Council.


\begin{references}
\bibitem{anisotropy} Only $g$ is involved and not $g^*$, as long as
the spin-anisotropy can be neglected.
\bibitem{why} Why is this material, which is similar to
$\lambda$-(BETS)$_2$FeCl$_4$, not insulating at zero field? This may
be because $J \sim 0.2 $ meV (see text) is smaller in this material
and the Ga material may not be as close to the Mott insulating phase.
The coexistence of AF and superconductivity can be justified by the
fact that the RKKY interaction near $Q_{AF}$ is not modified by the
occurrence of the superconducting state \cite{anderson}.
\end{references}
\end{document}